\documentclass[twocolumn,twocolappendix]{aastex63}

\usepackage[sort&compress]{natbib}
\usepackage{mathptmx}
\usepackage{helvet}
\usepackage{courier}
\usepackage[english]{babel}
\usepackage[T1]{fontenc}
\usepackage{color}
\usepackage{babel}
\usepackage{array}
\usepackage{amsmath}
\usepackage{graphicx}

\defcitealias{MalkMosk_2021}{Paper~I}

\accepted{May 15, 2022}

\submitjournal{ApJ}

\shorttitle{On the origin of observed cosmic ray spectrum below 100 TV}
\shortauthors{Malkov \& Moskalenko}

\begin{document}

\title{On the origin of observed cosmic ray spectrum below 100 TV}

\author[0000-0001-6360-1987]{Mikhail A. Malkov}
\email{mmalkov@ucsd.edu}
\affiliation{Department of Physics and CASS, University of California San Diego, La Jolla, CA 92093, USA}

\author[0000-0001-6141-458X]{Igor V. Moskalenko}
\email{imos@stanford.edu}
\affiliation{Hansen Experimental Physics Laboratory and Kavli Institute for Particle Astrophysics and Cosmology, Stanford University, Stanford, CA 94305, USA}

\begin{abstract}
Recent precise measurements of primary and secondary  cosmic rays (CRs) in the TV rigidity domain  have unveiled  a bump in their spectra, located between 0.5--50 TV.  We argue that a local shock may generate such a bump by increasing the rigidity of the preexisting CRs below 50 TV by a mere factor of  $\sim$$1.5$. Reaccelerated particles below $\sim$0.5 TV are convected with the interstellar medium (ISM) flow and do not reach the Sun, thus creating the bump. This single universal process is responsible for {\it the observed spectra of all CR species} in the rigidity range below 100 TV. We propose that one viable shock candidate  is the Epsilon Eridani star at 3.2 pc from the Sun, which is well-aligned with the direction of the local magnetic field. Other shocks, such as old supernova shells, may produce a similar effect. We provide a simple formula,  Eq.~(\ref{eq:FitForm}), that reproduces the spectra of all CR species with only two nonadjustable shock parameters, uniquely derived from the proton data. We show how our formalism predicts helium and carbon spectra and the B/C ratio.
\end{abstract}


\section{Introduction: the CR ``Bump''} 

The last decade was marked by discoveries of new features in the spectra of CR species. Unexpected breaks and variations in spectral
indices are being unveiled with high confidence in the GV--TV rigidity range that deemed as well-studied \citep{ATIC06, 2010ApJ...714L..89A, 2011Sci...332...69A, 2009BRASP..73..564P, 2011ApJ...728..122Y, 2014PhRvL.112o1103A, 2015PhRvL.114q1103A, 2015PhRvL.115u1101A, 2017PhRvL.119y1101A, 2018PhRvL.120b1101A, 2018PhRvL.121e1103A, PhysRevLett.124.211102, 2018JETPL.108....5A, 2019ARep...63...66A, 2019AdSpR..64.2546G, 2019PhRvL.122r1102A, eaax3793, Aguilar2021, 2019AdSpR..64.2559G}. These features bear the signatures of CR acceleration processes and their propagation history.

First discovered was a flattening above $R_{{\rm br}}'\approx0.5$ TV in rigidity in the CR proton and helium spectra \citep{ATIC06, 2010ApJ...714L..89A, 2011Sci...332...69A}, which propelled a slew of interpretations. However, with improved measurements of spectra of several primary ($p$, He, C, O, Ne, Mg, Si) and secondary\footnote{The mostly secondary species are those, which are rare in CR sources, but abundant in CRs. They are produced primarily in fragmentations of heavier species over their lifetime in the Galaxy.} (Li, Be, B) species by AMS-02 \citep{2015PhRvL.114q1103A, 2015PhRvL.115u1101A, 2017PhRvL.119y1101A, 2018PhRvL.120b1101A, 2018PhRvL.121e1103A, PhysRevLett.124.211102},
only two interpretations remain feasible. These are (i) the intrinsic
spectral break in the CR injection spectrum or populations of sources
with soft and hard spectra, a so-called ``injection scenario'' \citep{2012ApJ...752...68V},
and (ii) a break in the spectrum of interstellar turbulence translated
into the break in the diffusion coefficient, a so-called ``propagation
scenario'' \citep{2012ApJ...752...68V,2012PhRvL.109f1101B,AloisioKink_2015}.
The latter has very few free parameters and is well-supported \citep{2020ApJ...889..167B, Boschini_2020} by the
AMS-02 data.

A turning point was the recent discovery of a steepening in the proton and helium spectra at $R_{{\rm br}}''\sim$ 13 TV \citep{ATIC06,2018JETPL.108....5A, 2019ARep...63...66A, 2019AdSpR..64.2546G, 2019PhRvL.122r1102A, eaax3793}, see parameterizations in \citet{Boschini_2020}. Together with the
lower rigidity break $R_{{\rm br}}'$, the two features likely result
from a single physical process that generates a ``bump'' in the
CR spectrum between 0.5--50 TV. It peaks by a factor of 2.5--3
near $\approx$13 TV above the background CR spectrum extrapolated from low energies.

The discovery of the second break challenges conventional interpretations of the CR spectra outlined above. 
This finding also refutes earlier hypotheses of a contribution from a nearby CR accelerator, such as a supernova
remnant (SNR), that accelerates the bump particles out of the interstellar medium (ISM)
\cite[e.g.,][]{2020ApJ...903...69F, Fornieri2020, 2020FrPhy..1624501Y}, because
such scenario implies no breaks in the \textit{secondary} component
\citep{2012ApJ...752...68V}. Moreover, the unprecedented accuracy
with which the sharp break at $R_{{\rm br}}'$ in the CR spectrum is
measured severely constrains its remote or global origins as they would
lead to a smoother transition. The recently enhanced data accuracy even allows to identify 
the dominant turbulent regime through which these CRs propagate from the source to the observer. It was shown
to be of Iroshnikov-Kraichnan type (\citealt{MalkMosk_2021}, hereafter \citetalias{MalkMosk_2021}).

Another clue to the origin of the bump is the break in the spectra of secondary species that are also flattening
above the first break ($R_{{\rm br}}'$), similarly to the primaries,
albeit with a different spectral index. It implies that the secondary
species have already been present in the CR mixture that the bump
is made of. This leads us to conclude that \textit{(i) the bump has to be made out of the preexisting CRs with
all their primaries and secondaries that have spent millions of years
in the Galaxy.} Meanwhile, the sharpness of the breaks indicates that
\textit{(ii) the bump is formed locally, a short time before we observed
it.}

In this paper, we argue that a single universal process is responsible for the observed spectra of all CR species in the rigidity range below 100 TV. We provide a simple formula, Eq.~(\ref{eq:FitForm}), {\it that reproduces the spectra of all CR species with only two shock parameters derived from a fit to the proton data.} It then fully predicts the entire bump structure including two break rigidities ($R_{\rm br}'$, $R_{\rm br}''$), two respective jumps in the spectral index and two curvatures of the spectrum (sharpness) at the breaks--which would otherwise require six parameters for each CR species. 
We show how our formalism \citepalias{MalkMosk_2021} operates using the proton, helium, and carbon spectra, and the B/C ratio. 

\section{Local CR Reacceleration}\label{Reacc}

An immediate consequence of the above-mentioned reasoning is
that the bump particles are background CRs reaccelerated inside of the Local Bubble. 
Estimates show that prospective
sources should be within 3--10 pc of the Sun \citepalias{MalkMosk_2021}.
These sources can be of two kinds: point-like and extended. We argue
below that 
the difference between the two is significant. If
the
source is point-like, it also has an elongated magnetic flux tube.
The reaccelerated CRs propagate to the observer through it. Gradual
lateral losses from the flux tube and the Iroshnikov-Kraichnan 
\citep[ IK,][]{1964SvA.....7..566I, 1965PhFl....8.1385K}
MHD turbulence driven by the lateral pressure gradient of the reaccelerated
CRs critically reshape their rigidity (10-TV bump) and angular spectra
that fit to the most accurate recent data. By contrast, propagation
of CRs from an extended source through a broad channel without significant
lateral pressure gradient and IK-turbulence would require additional
assumptions to become consistent with those data.

The bow-shocks and stellar wind termination shocks (TS) of nearby passing
stars are obvious candidates of point-like sources. However, most
stars can be disqualified as they are not magnetically connected with
the heliosphere. Surprisingly, we found a remarkable alignment
of the local magnetic field with the Epsilon Eridani star at 3.2 pc
from the Sun \citepalias{MalkMosk_2021}. It also has an extensive
8000 au astrosphere and a thirty times higher than the solar mass
loss rate \citep{Wood2002}.

Nevertheless, simple estimates show that even a bow-shock of such
a significant size as $\sim$$10^{4}$ au might only marginally reaccelerate
preexisting CRs by the required 1.5 factor to generate the bump. On
the other hand, this can readily be done by a larger shock. Moreover,
as any local shock is likely to be weak, CR species with initially
steeper spectra experience a more significant reacceleration (see
Eq.~{[}\ref{eq:DSAsol}{]} in the  following
section), which is consistent
with observations of the secondaries. A particularly appealing source
candidate is a sizable imploding shock in the local interstellar medium,
whose existence is supported observationally \citep{2014A&A...567A..58G},
see discussion in Sect.~\ref{Sec:discussion}.

Notwithstanding the advantage of a large shock, this source is also
problematic, as we already mentioned. One serious contradiction comes
from the angular distribution of CRs in the bump rigidity range. It
rises sharply across the magnetic horizon and also has an enhanced
field-aligned component (see Fig.~11 in \citealt{Abeysekara2019} and
an extended discussion of available data in \citetalias{MalkMosk_2021}).
These angular anomalies can be produced simultaneously if a point-like source is located on
one side of the flux tube  connected to the solar system within several parsecs from the observer.
CRs that diffusively propagate from the source  along the flux tube
also leak through the tube's
lateral boundary. Therefore, not all the CRs that passed the observer will 
return. The observer  in the solar system will then see more CRs coming from the source than returning to it. The sharp contrast between
these CR groups on the intensity map is critically associated with the IK spectrum of the scattering turbulence.
The field-aligned component is not related to the CR losses but follows the solution of
the complete Fokker-Plank transport problem, not limited by the diffusive approximation.  It was shown \citep{malkov2017exact} that
the field-aligned component survives a longer propagation from the source  (up to 5-7 particle mean free paths) 
than might have been expected.

Despite the rise at the CR pitch-angle
of $90^{\circ}$ (magnetic horizon) and the field-aligned one-sided
enhancement have different physical causes, their coexistence 
have been demonstrated \citepalias{MalkMosk_2021} by solving
the proper CR propagation with lateral losses and the  IK
pitch-angle scattering. To our best knowledge, no other CR source
and transport regime configuration fits the current data in 1-100
TV range with dramatically shrunk statistical uncertainties \citep[][see also Fig.~\ref{fig:Proton-rigidity-spectrum} below]{Aguilar2021}.

The current data are so accurate and multifaceted (rigidity-, angular-,
and chemical composition spectra) that the ``reverse engineering''
of the 10-TV bump source and its propagation mode became possible
without identifying any particular object presumably responsible for
the bump. Still, the source is likely associated with a shock, whose
parameters can be constrained, but the shock may remain invisible
in the sky. In the Appendix, we provide a brief discussion of potentially
qualified shocks.

The situation with a well-described but unknown source is not entirely
satisfactory, so we pursue two parallel approaches to improve it.
One approach is to develop a ``generic''
shock reacceleration and CR propagation model along a magnetic flux
tube pointing to the solar system. It depends on an unknown shock
Mach number and its size-distance relation \citepalias{MalkMosk_2021}.
The model's key advantage is that these parameters are uniquely derived
from the best measured CR proton spectrum or its substitute. As the
latter, we use ``synthetic'' data provided
in \citet{Boschini_2020}, i.e., a physically motivated local interstellar
spectrum (LIS) of CR protons tuned to the data of different instruments
with adjustments for the appropriate heliospheric modulation. As the
slope and normalization of CR background LIS below the first break
are now accurately derived \citep{Boschini_2020}, the model takes
them as fixed input parameters. The turbulence index, found to be
$s=3/2$  (IK spectrum),
critically determines the proton spectrum, yielding the precise spectral
shape over four orders of magnitude in rigidity \citepalias{MalkMosk_2021}.
To further test our model, we turn to other species.

The second approach is to look at the Epsilon Eridani star as it is
magnetically connected with our Sun and its distance is also in the
right ballpark. However, as we noted, 
 its bow shock is somewhat weak,
making
the reacceleration of CR of several tenths of TV problematic. 
 This issue can be ameliorated considering that the bow shock and strong stellar wind TS coexist in the system, while
the resulting
bump spectrum is still primarily shaped during the CR propagation.
We provide  the corresponding formalism in the
Appendix. In the following section, we review and extend to a broader geometrical
setting a generic CR reacceleration and propagation model developed
in \citetalias{MalkMosk_2021}.  We now include the lateral particle losses in this model.

\section{CR Reacceleration and Propagation Model}\label{sec:Reaccel-and-Prop}

Only two parameters, constituting the bump amplitude, K, and rigidity,
$R_{0}$, suffice to predict the bump spectrum \citepalias{MalkMosk_2021}
without particle losses from the flux tube. The reaccelerated CR
density is elevated over the background by the factor 
\begin{equation}\label{eq:K}
K=\left(\gamma+2\right)/\left(q-\gamma\right),
\end{equation}
where $\gamma$ is the spectral index of the background CRs 
and $q>\gamma$ is that of the shock, $q=\left(r+2\right)/\left(r-1\right)$.
Here $r$ is the shock compression ratio, which we determine from
the proton fit. The enhanced spectrum appears as a bump at some distance
from the shock, as the reaccelerated CR with lower rigidity do not
reach the observer. There will also be a high-energy cutoff to the
CR enhancement, associated with the particle losses from the flux
tube and the limited shock size, which we also discuss later.

The background CR spectrum is $f_{\infty}\propto R^{-\gamma},$ whereas
the shock would accelerate the freshly injected thermal particles
to $R^{-q}$ spectrum. However, as the shock is likely to be weak,
that is $q>\gamma\approx2.85$ (for protons), the freshly accelerated
CRs are submerged into the reaccelerated CR component. The CR density
measured at a distance $z$ from the shock is (\citealt{BlandOst78}, \citetalias{MalkMosk_2021}):
\begin{equation}
f\left(R,z\right) = f_{\infty}\left(R\right)\left[1+Ke^{-\Phi\left(R,z\right)}\right],\label{eq:DSAsol}
\end{equation}
where $K$ is defined in Eq.~(\ref{eq:K}).
The flux suppression exponent, $\Phi$, is a path integral of particles
propagating along the magnetic flux tube from the shock. One may distinguish
between two cases of the relative motion of the shock and the observer,
although the rigidity dependence of $\Phi\left(R\right)$ is similar in both
cases.

\subsection{Propagation from quasi-parallel portions of shock}

If the shock normal makes an angle $\vartheta_{Bn}$ with respect
to the local magnetic field not too close to $\pi/2$, the CR distribution
along the flux tube is determined by the balance between the CR convection
with the flow and diffusion against it. In this case we have 
\begin{equation}
\Phi=u\int_{0}^{z}\frac{dz}{\kappa_{\parallel}\left(R,z\right)}+\mathcal{L}\left(R\right).\label{eq:FiDef}
\end{equation}
Here $u=u_{sh}/\cos\vartheta_{Bn}$ with $u_{sh}$ being the shock
velocity, and $\kappa_{\parallel}$ is the parallel component of the
diffusion tensor. The term $\mathcal{L}\left(R\right)$ describes
lateral losses from the flux tube. The bump characteristic rigidity
$R_{0}$ is introduced by the relation $\Phi\approx$$\left(R_{0}/R\right)^{1/2}+\mathcal{L}$.
Here the power $1/2$ corresponds to the  IK
turbulence. The parameter $R_{0}$ combines the distance to
the shock, $z$, velocity $u$ and the turbulence power, that defines
$\kappa_{\parallel}$ in Eq.~(\ref{eq:FiDef}). According to \citetalias{MalkMosk_2021},
$R_{0}\propto u^{2}z^{2}/l_{\perp}$ where the characteristic scale
of the CR pressure gradient across the flux tube is $l_{\perp}^{-1}=P_{\text{CR}}^{-1}$$\partial P_{\text{CR}}$/$\partial r$.
It sets the amplitude of magnetic perturbations driven by this gradient.

\subsection{Propagation from quasi-perpendicular portions of shock}

In the case $\vartheta_{Bn}\approx\pi/2$, satisfied at some fraction
of almost any curved shock, we use the balance between the divergence
of convective and diffusive CR fluxes in the shock frame, $\boldsymbol{v}\cdot\nabla F\sim\nabla_{\parallel}\kappa_{\parallel}\nabla_{\parallel}F$,
which is consistent with Eqs.~(\ref{eq:DSAsol})--(\ref{eq:FiDef}).
Here $F$ is the full-bump distribution related to $f$ below. The
CR diffusion proceeds primarily along the field, but the convective
term is $\boldsymbol{v}\cdot\nabla F\approx\boldsymbol{v}_{\perp}\cdot\nabla_{\perp}F$,
which we may write assuming the ISM wind being in $y$-direction and
magnetic field in $z$-direction. So we have v$\partial F/\partial y\approx\kappa_{\parallel}\partial^{2}F/\partial z^{2}$,
assuming that $\kappa_{\parallel}$ is independent of $z$, for a
simple illustration. The result will then have a similar to the former
case exponential dependence, 
\begin{equation}
\propto\exp\left[-vz^{2}/4y\kappa_{\parallel}\left(R\right)\right].\label{eq:near-zone2}
\end{equation}
Averaging over $y$, or rather over radial coordinate inside the tube,
$r_{\perp}$, depends on the shock structure and CR propagation in
the tube that controls the losses. 

\subsection{Lateral Losses of CRs}

Lateral losses from the flux tube affect high-rigidity end of the
10-TV bump, roughly between 10-50 TV. Our goal is to include them
in a way applicable to different CR propagation regimes. Assuming
a quasi-stationary distribution of reaccelerated CRs around a moving
reaccelerator (e.g., a stellar bow-shock) we balance the following
channels of particle transport: convection along the field, two components
of diffusion (parallel and perpendicular to the B-field), and the lateral
losses.

If the convective transport is significant, it is convenient to use
the bump component $f_{\text{b}}\left(R,z\right)\equiv f-f_{\infty}$
in Eq.~(\ref{eq:DSAsol}) as a radially averaged part of its full
distribution, $F$. Namely, $f_{\text{b}}=2a_{\perp}^{-2}\int_{0}^{a_{\perp}}F\left(R,z,r_{\perp}\right)r_{\perp}dr_{\perp}$,
where $a_{\perp}$ is the tube's radius. The CR flux at $r_{\perp}=a_{\perp}$
is $-2\pi\kappa_{\perp}\partial F/\partial r_{\perp}|_{r_{\perp}=a_{\perp}}.$
Here $\kappa_{\perp}$ is a cross-field component of the CR diffusion
tensor. Particle losses from the flux tube depend on the distance
from the source because the self-driven turbulence decreases with
it. We use the following general relation between the two components
of $\kappa$, \citep[e.g.,][]{Drury83}, $\kappa_{\perp}\approx\kappa_{B}^{2}/\kappa_{\parallel}\approx\kappa_{B}k\delta B_{k}^{2}/B^{2}$,
where $\kappa_{B}=cr_{g}/3$ is the Bohm diffusion, $r_{g}$ is the
Larmor radius, and $\delta B_{k}^{2}\ll B^{2}/k$ is the spectral
energy density of Alfven waves evaluated at the resonant wave number
$k=r_{g}^{-1}$. Alfvenic fluctuations inside the magnetic flux
tube filled with reaccelerated CRs were shown to obey the Iroshnikov-Kraichnan
law, $\delta B_{k}^{2}\propto k^{-3/2}$ \citepalias{MalkMosk_2021}. In
this case, $\kappa_{\perp}\propto R^{3/2}$, while $\kappa_{\parallel}\propto R^{1/2}$.
It follows that for large $R$ the path integral $\Phi$ in Eq.~(\ref{eq:FiDef})
will be dominated by the second term, associated with lateral particle
losses from the tube rather than their propagation along the field
(the first term).

If the CR source velocity has a sizable component along the B-field
the lateral losses are balanced by the CR convection along the field.
The diffusion in that direction is suppressed because of the self-driven
IK-turbulence. The CR flux from the magnetic tube, $-2\pi\kappa_{\perp}\partial F/\partial r_{\perp}|_{r_{\perp}=a_{\perp}}$,
can be estimated by recognizing that the fluctuation level must drop
at the tube's edge, $r_{\perp}=a_{\perp}$, as these fluctuations
are driven by the excessive pressure of reaccelerated CRs inside the
tube. Beyond this radius, CRs diffuse primarily along the field, since
the level of the background ISM fluctuations is typically very small,
$k\delta B_{k}^{2}/B^{2}\sim10^{-5}$ at a scale $\sim10^{12}-10^{13}$cm
\citep[e.g.,][]{AchterbergBR94}. Hence, the scattering events resulting
in the displacement of the particle guiding center are increasingly
rare. 

Upon reaching the edge of the flux tube particles almost instantaneously
escape along the field, since $\kappa_{\perp}\sim k^{2}\left(\delta B_{k}/B\right)^{4}\kappa_{\parallel}\ll\kappa_{\parallel}$,
notwithstanding the tube's length being much larger than its lateral
size. Under these conditions, the most plausible estimate for $\partial F/\partial r_{\perp}|_{r_{\perp}=a_{\perp}}$
is $\sim-F/r_{g}$. The CR flux through the lateral boundary of the
tube scales with rigidity as $\sim$$\kappa_{\perp}F/r_{g}\propto R^{1/2}F.$
An equilibration of convective CR transport $\boldsymbol{v}\cdot\nabla F$
with these losses yields the following approximate relation $v_{z}\partial f_{b}/\partial z\approx-\kappa_{\perp}f_{b}/r_{g}a_{\perp}$.
Here $v_{z}$ is the projection of the CR source velocity on the field
direction. We thus obtain the following exponential decay of the CR
density along the tube: $f_{b}\propto\exp\left[-\sqrt{R/R_{L}}\right]$.
Here $R_{L}$$\propto1/z^{2}$ is the rigidity cut-off associated
with the lateral losses and $z$ is the current distance to the source.
The cut-off $R_{L}$ also depends on the flow geometry and details
of the CR escape that we discussed using only order of magnitude estimates
above. Therefore, the path integral in Eq.~(\ref{eq:FiDef}) at fixed
$z$ can be represented as 
\begin{equation}
\Phi=\left(R_{0}/R\right)^{1/2}+\left(R/R_{L}\right)^{1/2}.\label{eq:FiTot}
\end{equation}

Fortunately, the same form of high-rigidity cut-off is applicable
when the convective transport is negligible because of the flow geometry,
i.e., $v_{z}\approx0$. In this case the lateral CR diffusion is balanced
by that along the field:

\begin{equation}\label{eq:par-perp-diff-bal}
\frac{\partial}{\partial z}\kappa_{\parallel}\frac{\partial F}{\partial z}+\frac{1}{r_{\perp}}\frac{\partial}{\partial r_{\perp}}r_{\perp}\kappa_{\perp}\frac{\partial F}{\partial r_{\perp}}=0.
\end{equation}
After substituting $\kappa_{\perp}=\kappa_{B}^{2}/\kappa_{\parallel}$,
assuming that $\kappa_{\parallel}$ is separable, i.e., $\kappa_{\parallel}=\kappa_{0}\left(R\right)\kappa_1\left(r_{\perp}\right)P\left(z\right)$,
and introducing a new variable $\zeta=\int_{0}^{z}dz/P$, the last
equation rewrites as follows

\begin{equation}
\frac{\kappa_{0}^{2}}{\kappa_{B}^{2}}\frac{\partial^{2}F}{\partial\zeta^{2}}+\frac{1}{r_{\perp}\kappa_1}\frac{\partial}{\partial r_{\perp}}\frac{r_{\perp}}{\kappa_1}\frac{\partial F}{\partial r_{\perp}}=0\label{eq:dif-bal-2}.
\end{equation}
This equation can be solved by separating the variables $F=\Psi\left(\zeta\right)\phi\left(r_{\perp}\right)$:

\[
\frac{\kappa_{0}^{2}}{\kappa_{B}^{2}}\frac{1}{\Psi}\frac{\partial^{2}\Psi}{\partial\zeta^{2}}=-\frac{1}{r_{\perp}\kappa_1\phi}\frac{\partial}{\partial r_{\perp}}\frac{r_{\perp}}{\kappa_1}\frac{\partial\phi}{\partial r_{\perp}}=\lambda^{2},
\]
where $\lambda^{2}$ is a separation constant, which is also the spectral
parameter of the problem for $\phi$ defined by the boundary conditions
$\phi\left(r_{\perp}=a_{\perp}\right)=0$ and the requirement of regularity
of $\phi$ at $r_{\perp}=0$ imposed on the second equation above.
For $\kappa_1=const$, its solution is given by Bessel function, $J_{0}$,
$\phi_{n}\left(r_{\perp}\right)=J_{0}\left(\mu{}_{n}r_{\perp}/a_{\perp}\right)$,
where $\mu_{n}$ is its n-th root, $J_{0}\left(\mu_{n}\right)=0$,
$\mu_{1}\approx2.5$, $\mu_{2}\approx5.5,\dots$ In this case, the
spectrum is $\lambda_{n}=\mu_{n}/a_{\perp}\kappa_1$. For $\kappa_1\left(r_{\perp}\right)\neq const$,
the eigenvalues $\left\{ \lambda_{n}\right\} _{n=1}^{\infty}$ also
grow rapidly with $n$. Therefore, expanding the solution of Eq.~(\ref{eq:dif-bal-2})
in a series of eigenfunctions $\left\{ \phi_{n}\right\} _{n=1}^{\infty}$
with coefficients $\Psi_{n}\left(\zeta,R\right)$,

\[
\Psi_{n}=C_{n}\left(R\right)\exp\left\{ -\lambda_{n}\frac{\kappa_{B}}{\kappa_{0}}\int_{0}^{z}dz/P\right\} ,
\]
we can retain only the first term in the series at the observer position
$z$. Note, that particle rigidity, $R$, enters Eq.~(\ref{eq:par-perp-diff-bal})
as a parameter, so that the solution can be represented as follows

\begin{equation}
F=\sum_{n=1}^{\infty}\Psi_{n}\left(\zeta,R\right)\phi_{n}\left(r_{\perp}\right)\approx C_{1}\phi_{1}\exp\left\{ -\lambda_{1}\frac{\kappa_{B}}{\kappa_{0}}\int_{0}^{z}\frac{dz}{P}\right\} .\label{eq:dif-bal-sol}
\end{equation}
The prefactor $C_{1}\left(R\right)\phi_{1}\left(r_{\perp}\right)$
is determined by the spectrum generated by the reaccelerator and the
near-zone propagation. Considering the CR reacceleration at a generic
shock wave, for example, $C_{1}$ includes the background CR spectrum
($f_{\infty}$ in Eq.~{[}\ref{eq:DSAsol}{]}) and the first term in
Eq.~(\ref{eq:FiDef}) or Eq.~(\ref{eq:near-zone2}). The choice between
the two depends on the accelerator's motion relative to the observer
and the ambient B-field, as considered earlier in this section. In
either case, the exponential dependence on rigidity is in the form
given by the first term in Eq.~(\ref{eq:FiTot}). Nevertheless, different
expressions for the CR bump parameter $R_{0}$ apply, as considered
in \citetalias{MalkMosk_2021} in more detail. The significance of the
result in Eq.~(\ref{eq:dif-bal-sol}) is that the exponential rigidity
cut-off has the same form as in Eq.~(\ref{eq:FiTot}), $\exp\left(-\sqrt{R/R_{L}}\right)$.
This follows from the relation $\kappa_{B}/\kappa_{0}\propto\sqrt{R}$
that we have discussed earlier.

To conclude this section, the CR bump-defining factor, $\exp\left[-\Phi\left(R\right)\right]$,
is generically given by Eq.~(\ref{eq:FiTot}) with two parameters $R_{0}$
and $R_{L}$ that can be determined analytically (see above and \citetalias{MalkMosk_2021}
for detailed calculations), depending on the particular reacceleration scenario. Alternatively,
they can be extracted from fitting this formula to the data (next
section). As this factor bears more on propagation and losses than
reacceleration, it can also be applied to the case of CR reacceleration
at a stellar TS, which we consider in detail in the Appendix.
It boosts the stellar bow-shock reacceleration, thus easing the observational
constraints on the observed CR bump.

\begin{deluxetable*}{ccccccc}[tb!]
\tablecaption{Model parameters and fit results for the proton spectrum.\label{tab:Model-parameters-and}}
\tablewidth{0pt}
\tablehead{
\colhead{Parameter (St.~err.~\%)} & \colhead{$R_{0}$(GV)} & \colhead{$R_{L}$(GV)} & \colhead{$q$} &
\colhead{$K=\left(\gamma+2\right)/\left(q-\gamma\right)$}  & \colhead{$\chi_{\text{min}}^{2}$/dof} & \colhead{dof}
}
\startdata
Realistic Model (RM) & 5878 (3.5\%) & $2.24\times10^5$ (28\%) & 4.2 & 3.59 (4.9\%) & 0.10 & 76-3\\ 
Loss-Free Model (LF) & 4795 (3.2\%) & $\infty$ & 4.7 & 2.58 (2.9\%) & 0.19 & 76-2\\
\enddata
\end{deluxetable*}

\begin{deluxetable}{ccccc}[tb!]
\caption{Input parameters for CR species derived from their LIS \citep{Boschini_2020}.\label{tab:Input-parameters-and}}
\tablewidth{0pt}
\tablehead{
\colhead{Parameters} & \colhead{protons} & \colhead{helium} & \colhead{boron} & \colhead{carbon}
}
\startdata
$A_s$ (m$^{-2}$ s$^{-1}$ sr$^{-1}$ GV$^{-1}$) & $2.32\times10^{4}$ & 3410 & 79 & 109 \tabularnewline
$\gamma_s$ & 2.85 & 2.76 & 3.1 & 2.76 \tabularnewline
\enddata
\end{deluxetable}

\section{Probing the Shock with Reaccelerated Protons}

Following the formalism developed in Paper I
and the results of the preceding section, we represent the spectrum
of an arbitrary CR species in the following form, Eqs.~(\ref{eq:DSAsol}),
(\ref{eq:FiTot}): 
\begin{equation}\label{eq:FitForm}
 f_{s}\left(R\right)=A_{s}R^{-\gamma_{s}}\left\{ 1+\frac{\gamma_{s}+2}{q-\gamma_{s}}\exp\left[-\sqrt{\frac{R_{0}}{R}}-\sqrt{\frac{R}{R_{L}}}\right]\right\},
\end{equation}
where $A_{s}$ and $\gamma_{s}$ are the normalization and spectral
index of CR species ($s=p$, He, B, C, \dots) that will be fixed using the data at $R\ll R_{0}$
(the so-called background CRs). 
These data are not related to the bump phenomenon described by the
remaining three parameters: $q$, $R_{0}$, and $R_{L}$, where,
$q=\left(r+2\right)/\left(r-1\right)>\gamma_{s}$. 

In the first step we remotely sense the shock parameters using the
precision measurements of the local proton spectrum. The best available
data in the GV--TV domain are provided by AMS-02, CALET, and DAMPE \citep{Aguilar2021, 2019PhRvL.122r1102A, eaax3793}. To mitigate the systematic discrepancies between these instruments we use overlaps of their rigidity ranges as follows.

\begin{figure}[tb!]
\includegraphics[width=1\linewidth]{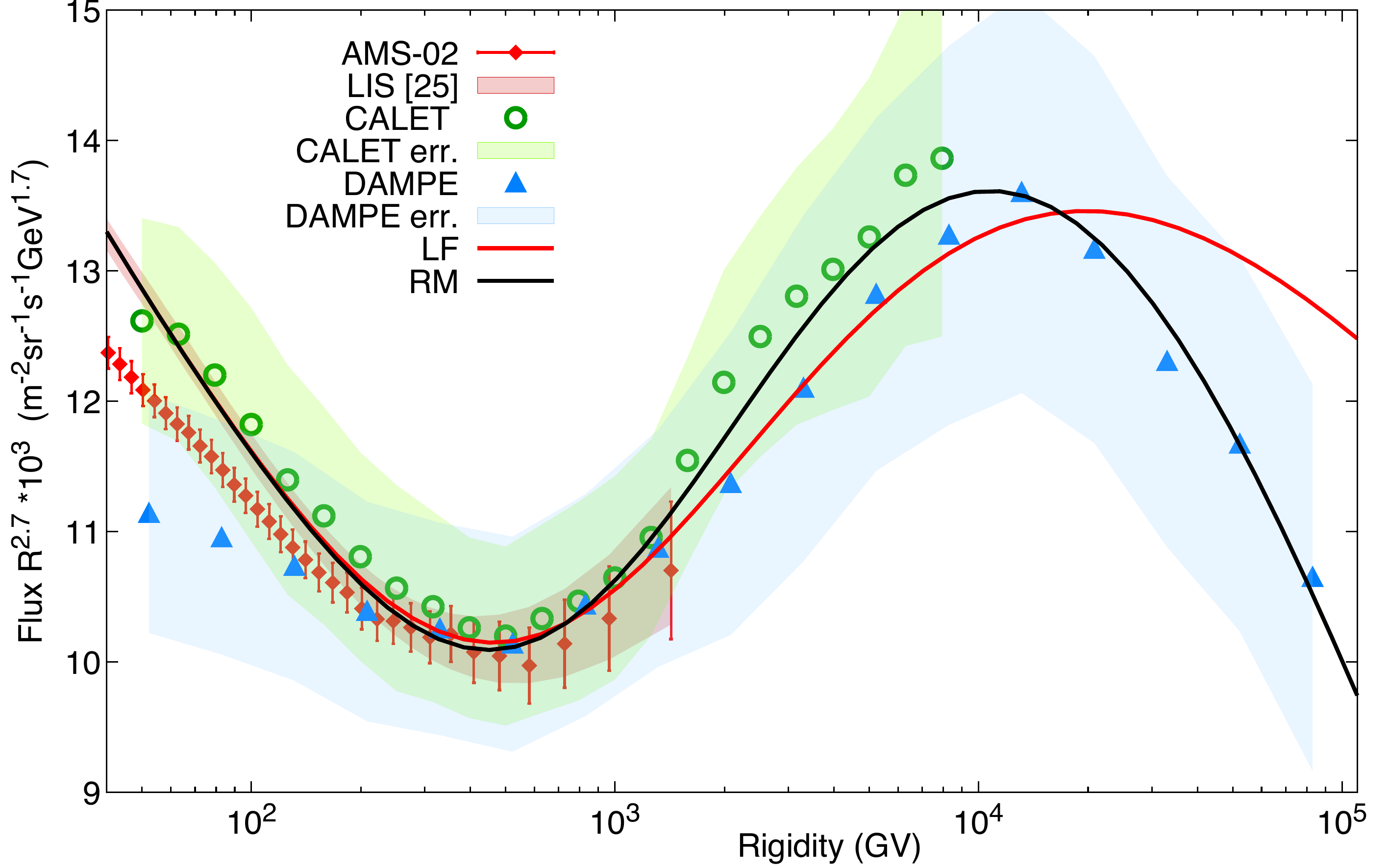}
\caption{Proton rigidity spectrum from AMS-02, CALET,
and DAMPE \citep{Aguilar2021, 2019PhRvL.122r1102A, eaax3793}. CALET and DAMPE data were modified to
adjust for systematic differences between the three instruments, see
text for details. The curves show the model calculations, Eq.~(\ref{eq:FitForm}),
for the cases of $R_{L}=\infty$ (LF model) and a more
realistic for high rigidities case of finite losses $R_{L}<\infty$
(RM model). The model and input parameters are summarized in Tables~\ref{tab:Model-parameters-and},
\ref{tab:Input-parameters-and}. \label{fig:Proton-rigidity-spectrum}}
\end{figure}

The most accurate measurements are provided by AMS-02, which has several
independent systems that allow for data cross checks. To
eliminate the solar modulation effects below the first break rigidity, we
use the proton LIS \citep{Boschini_2020}, which was derived
using the AMS-02 data. The CALET and DAMPE data are then modified to normalize their first break rigidity $R_{{\rm br}}'$
and their flux at $R_{{\rm br}}'$ (minimum in the spectrum in Fig.~\ref{fig:Proton-rigidity-spectrum})
to the values derived from the proton LIS. This was achieved by applying
the following transformations: $R=\tilde{R}/1.12$, Flux$=\widetilde{{\rm Flux}}\times$1.09
to CALET, and $R=\tilde{R}/0.95$, Flux$=\widetilde{{\rm Flux}}\times$0.98
to DAMPE data. Here $\tilde{R}$ and $\widetilde{{\rm Flux}}$ are
the published rigidity and flux values \citep{2019PhRvL.122r1102A, eaax3793}.
These adjustments are well within the instrumental error bars. The
model parameters derived from the best fit to the modified data of
all three instruments are summarized in Table~\ref{tab:Model-parameters-and}.

The LIS spectra below the break were used to obtain the model input parameters listed in Table~\ref{tab:Input-parameters-and}.
Fig.~\ref{fig:Proton-rigidity-spectrum} shows the
proton data along with two models, where we have considered the cases of loss-free (LF) model $R_{L}=\infty$
and realistic model (RM) $R_{L}<\infty$, which is more accurate
for $R\gtrsim R_{{\rm br}}''$. 
The RM model provides a better match of the data above 10 TV indicating that the shock size or other geometry constraints are at play. \textit{Regardless of their nature, all model parameters, $q$, $R_{0}$, $R_{L}$, are now fixed.} 

\section{Calculating Spectra of Other Elements}

\begin{figure}[tb!]
\includegraphics[width=1\linewidth]{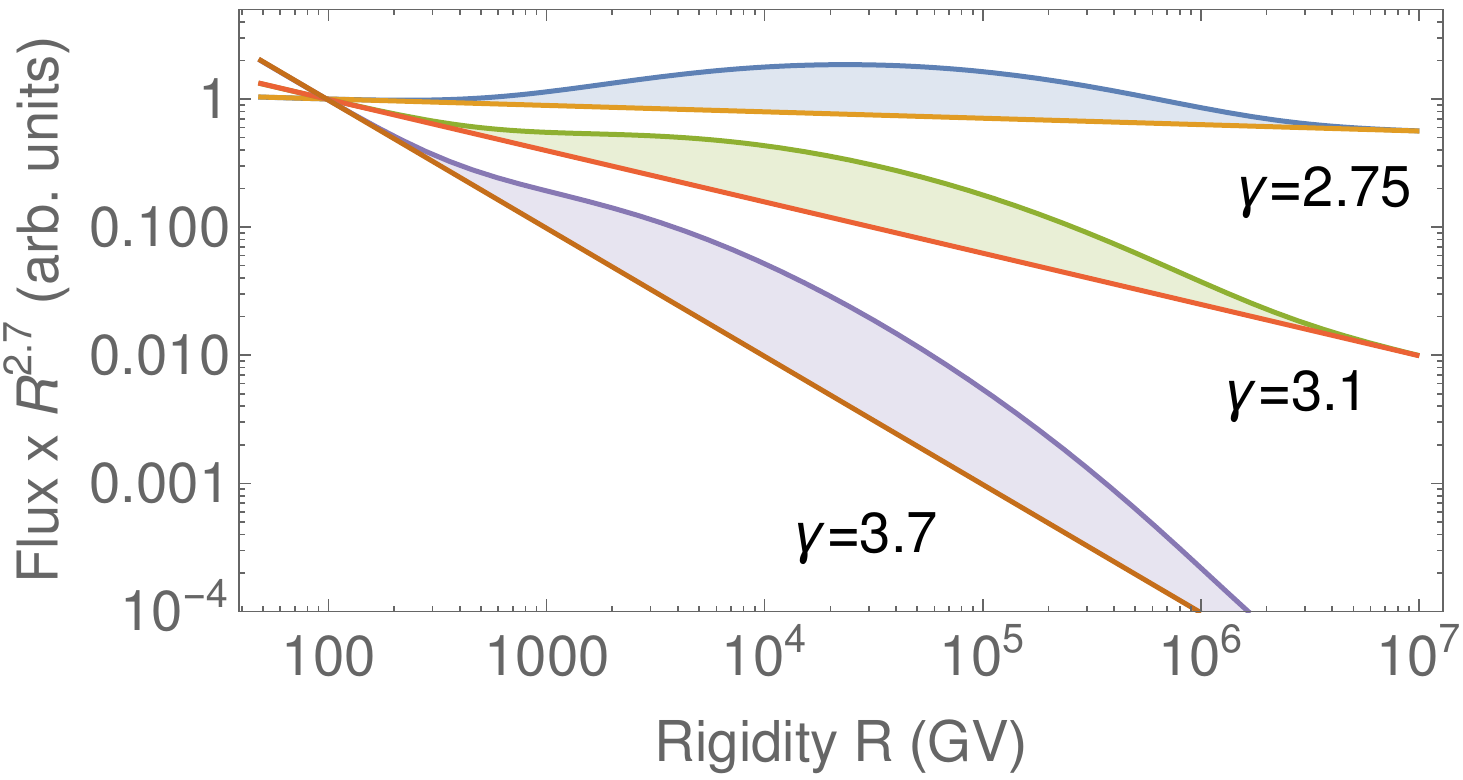}
\caption{The bump strengths in the CR spectra with different spectral indices
corresponding to the helium and boron LIS (Table~\ref{tab:Input-parameters-and}),
and $\gamma_{{\rm arb}}=3.7$ (for illustration),
calculated using Eq.~(\ref{eq:FitForm}) and compared to the straight
background power-laws. The ratios of the enhanced fluxes at $R=30$
TV to the background values for the three different values of $\gamma$
are 2.46, 3.06, and 6.05, respectively. \label{fig:Bump-strength} }
\end{figure}

According to Eq.~(\ref{eq:FitForm}), the species are differentiated solely by their spectral index $\gamma_{s}$ below the break $R_{{\rm br}}'$. We illustrate this in Fig.~\ref{fig:Bump-strength}, where we use the spectral indices corresponding to the helium and boron LIS, and an arbitrary index corresponding to a steeper spectrum, $\gamma_{{\rm arb}}=3.7$. One can see that CRs with steeper background spectra undergo stronger reacceleration.

\begin{figure}[tb!]
\includegraphics[width=1\linewidth]{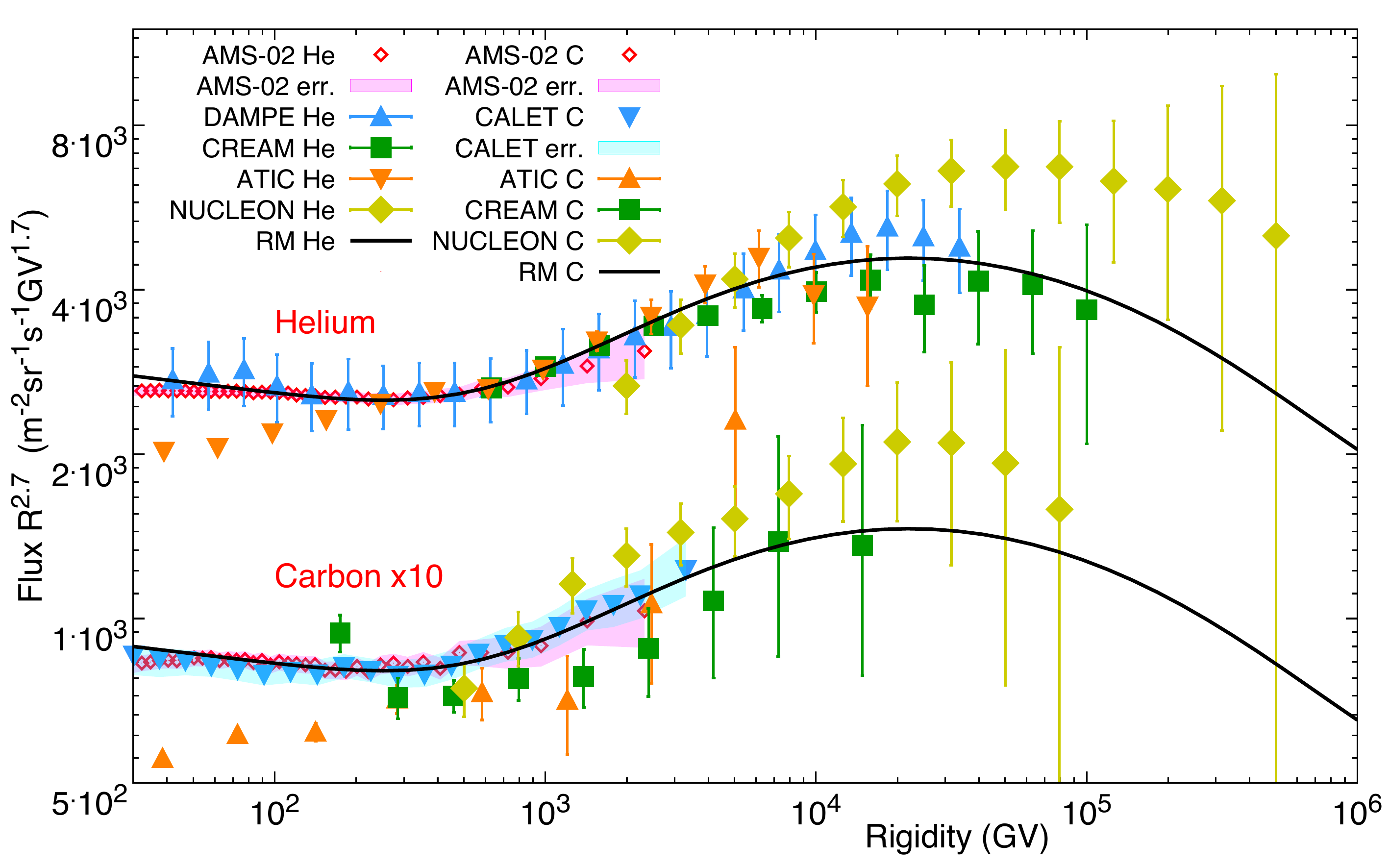}
\caption{The calculated helium and carbon spectra are compared with available helium data from AMS-02, DAMPE, CREAM, ATIC, and NUCLEON \citep{Aguilar2021, 2021PhRvL.126t1102A, 2020JETPL.111..363K, 2009BRASP..73..564P, 2017ApJ...839....5Y} and carbon data from AMS-02, CALET (multiplied by 1.27), ATIC, CREAM, and NUCLEON \citep{2009ApJ...707..593A, 2020PhRvL.125y1102A, Aguilar2021, 2020PhLB..81135851K}. The RM model calculations are using Eq.~(\ref{eq:FitForm}) with parameters provided in Tables~\ref{tab:Model-parameters-and}, \ref{tab:Input-parameters-and}. \label{fig:He-C-plot}}
\end{figure}

Fig.~\ref{fig:He-C-plot} shows the primary helium and carbon spectra. The DAMPE and CREAM helium spectra were normalized to the helium LIS by applying factors 1.08 and 1.05, correspondingly. Following \citet{2020PhRvL.125y1102A} the CALET spectrum of carbon was multiplied by 1.27 to agree with AMS-02 data. Given the same spectral indices below the first break $R_{{\rm br}}'$, the shapes of helium and carbon spectra are identical.

\begin{figure}[tb!]
\includegraphics[width=1\linewidth]{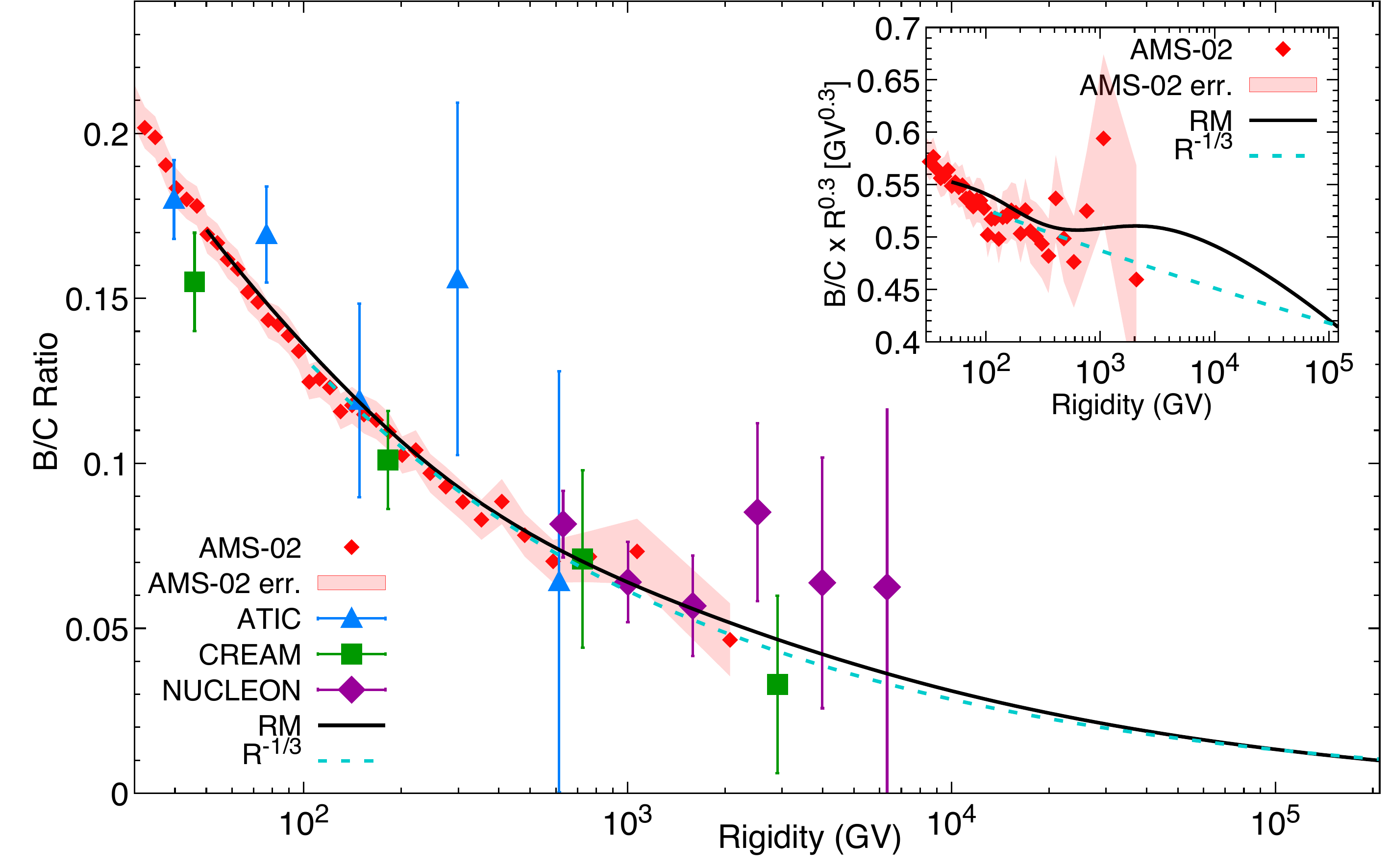}
\caption{The calculated B/C ratio is compared with AMS-02, CREAM, ATIC, and NUCLEON data \citep{Aguilar2021, 2008APh....30..133A, 2019AdSpR..64.2559G, 2008ICRC....2....3P}. The RM model calculations are using Eq.~(\ref{eq:FitForm}) with  parameters provided in Tables~\ref{tab:Model-parameters-and}, \ref{tab:Input-parameters-and}.\label{fig:B/C-ratio}}
\end{figure}

We are also comparing our model calculations with the measurements of the B/C ratio that is widely used to evaluate the 
CR residence time in the Galaxy. It is
measured in a wider rigidity range and with higher precision than the individual spectra of boron and carbon.
Fig.~\ref{fig:B/C-ratio} shows the calculated B/C ratio compared with available data. The inset shows the ratio multiplied by a factor $R^{0.3}$ to emphasize the fine details in the bump range. 
Our model shows $\approx$10\% deviation from predictions for the Kolmogorov turbulence $\propto R^{-0.33}$ \citep{1941DoSSR..30..301K}  in the bump range. 

\section{Background spectra}

Here we comment on the origin of the background spectra of different CR species below $R_{{\rm br}}'$. A physical explanation
for the observed difference between $\gamma_{p}$ and $\gamma_{s}$
of primary elements with effective mass to ion-charge ratios $\left\langle A/Q\right\rangle $
$\approx2$, or higher has already been suggested \citep{m98,MDSPamela12,Hanusch2019ApJ}.
They develop harder than proton spectra (the index difference $\gamma_{p}-\gamma_{s}\approx0.1$) because SNR shocks extract
them from thermal plasma most efficiently when these shocks are at
their strongest. The time-integrated spectra of heavier elements become
then harder than that of the protons.

This bias stems from the shock related turbulence whose amplitude increases with the Mach number. Resonantly driven by protons, it curbs them tighter downstream than heavier nuclei, so the latter may cross the shock to gain energy. As the above differentiation mechanism depends exclusively on $\left\langle A/Q\right\rangle $,
its salient consequence is that all elements with the same $\left\langle A/Q\right\rangle $
must have the same ultrarelativistic rigidity spectra \citep{Malkov2018a}.
No exceptions to this rule have been reported so far \citep{Aguilar2021}.

The steep background spectrum of the secondaries is related to a decreasing
residence time of CR species in the Galaxy as rigidity increases.
The secondaries are already produced in fragmentations of primary
CRs with a steep spectrum that is further steepened due to the rigidity-dependent
residence time.

\section{Discussion}\label{Sec:discussion}

There is no reason to believe that the observed bump in the CR spectrum
is unique. Similar bumps might be generated in other Galactic locations
where shocks are present. Whether their integral contribution competes
favorably with the CR reacceleration by the ISM turbulence is too
early to say. Theoretical estimates of the turbulent reacceleration
\citep{1994ApJ...431..705S, 2006ApJ...642..902P, DruryStrong2017}
suggest that it supplies up to 50\% of the power required to maintain
the CRs against losses. However, this estimate is difficult to confirm
using observations, as the reaccelerated CRs are inseparable from
the background.

By contrast, a contribution of CR-reaccelerating shock is identifiable
and, as we argued, is currently observed. The next step is to create
an inventory of such shocks and integrate their contributions. Our
results show that such shocks can temporarily modify the local CR
spectrum. Studies of time-dependent CR transport show similar modifications
\citep{2019ApJ...887..250P}. However, the significance of our result
is that we do not need such a major event as a supernova explosion
nearby to modify the local CR spectrum.

The local interpretation of the 10-TV bump has broader consequences
in terms of our understanding of the CR spectrum. First, the B/C ratio
is also affected by such modifications. Hence, the observed B/C ratio
cannot be used to estimate the CR confinement time without corrections
for the reacceleration effects due to the passing stars and other
shocks in the local interstellar medium. We speculate that even the
well-known ``knee'' at $\sim$3
PV could also be interpreted as a ``big bump''
in the following fashion. The spectrum hardens below the knee due
to a reacceleration by some shock, which is consistent with observations.
How exactly it hardens is not clear since these observations are not
nearly as accurate as the new observations in the region \ensuremath{\lesssim}
10 TV discussed in this paper. The ``local''
shock that might be responsible for the reacceleration needs to be
2-3 orders of magnitude larger, e.g., a spiral density shock \citep{1972ApJ...173..557S}.
According to Eq.~(\ref{eq:DSAsol}), weaker shocks substantially affect
steep background CR spectra. The knee itself is then generated by
physical processes similar to those responsible for the 10-TV bump
formation addressed earlier.

Returning to the subject of this paper, let us discuss possible local
CR reaccelerators. Apparently, the reacceleration power requirements
are modest because the background CR spectrum is steep ($\propto R^{-2.85}$).
Its slight upshift (by a factor $\approx1.36)$ in rigidity suffices for    the bump to be formed.
However, its fine structure depends on 
the self-generated turbulence \citepalias{MalkMosk_2021}, through which the reaccelerated CRs propagate
to the observer along the magnetic field.
The low-energy reaccelerated particles ($R\lesssim1$
TV) do not reach the observer as they are convected downstream with
the ISM flow. This low-energy screening makes the distance to the
object a defining parameter for the bump.

A more detailed formalism
in \citetalias{MalkMosk_2021} shows that the observed bump is a factor
2.4 enhancement over the background spectrum at its maximum near 10
TV (after the spectrum is multiplied by $R^{2.7}$). Even a weak shock
of a size comparable with the particle Larmor orbit can provide the
required upshift, operating in an SDA regime (shock-drift acceleration).
All it takes is that the shock overruns a significant portion of the
particle Larmor orbit before the particle escapes the shock front
along the field line. The full orbit crossing-time is $r_{g}/U_{\text{shock}}\approx$30
yrs for $r_{g}\sim3\times10^{15}$ cm (a 10-TV proton in a 10 $\mu$G
field \footnote{The direct measurements of Voyager 1, 2 give 7-8 $\mu$G just outside
of the heliosphere \citep{Burlaga2019}.}) and $U_{\text{shock}}\approx30$ km sec$^{-1}$. The flow near the
shock front is likely turbulent, so the particle confinement in the
field direction can be taken at the Bohm level. Particles do not diffuse
beyond $\sim r_{g}\sqrt{c/U_{\text{shock}}}$ from the shock during
the orbit crossing.

While these numbers appear restrictive for the Epsilon Eridani bow
shock, which is 8000 au $\sim10^{17}$ cm, it can still cross orbits
of 10-TV particles before they diffuse away along the field. The bow
shock tail may even exceed $10^4$ au. Because the quasi-one-dimensional
diffusion of particles along the field is recurrent, they will have
a fair chance to interact with the shock repeatedly and climb up its
flanks to the apex, thus gaining more energy via the SDA process.
Stronger nondiffusive confinement of particles to a bow shock near
its head is also possible by modifying the shock magnetic environment
by energetic particles themselves. For example, such phenomena are
observed at the Earth's bow shock \citep[magnetic cavities, hot flow anomalies, etc.,][]{Sibeck2021}.
In addition, the diamagnetically expelled field ahead of the shock
may create a magnetic trap for the particles, thus extending their
interaction with the shock (see Appendix). The particle energy gain
after a single Larmor-circle shock crossing can be inferred from the
conservation of the particle magnetic moment: $p_{\perp}^{2}/B=const$,
where $p_{\perp}$ is the perpendicular to the B-field component of
particle momentum. Therefore, even a factor of two magnetic field
compression would suffice to make the bump visible.

The above constraints on the energy gain may be further relaxed in
the case of a powerful stellar wind TS (e.g., Epsilon
Eridani star), as the shock speed is likely to exceed the bow-shock
velocity by two orders of magnitude (see Appendix). In addition, the
Cranfield-Axford effect \citep{1972NASSP.308..609A} may substantially
increase the magnetic field in the surrounding area filled by the
shocked stellar wind. Analytic models for the CR spectrum re-power
by the stellar terminal shocks are available \citep[e.g.,][]{Webb1985},
showing a relatively high, typically 40\% efficiency of the wind energy
conversion into the reaccelerated CRs. Observations also indicate that 
powerful stellar winds accelerate particles efficiently \citep{Rangelov_2019}.

Shocks that are significantly larger than stellar bow-shocks are almost
certainly present in the solar system proximity and can remove the
maximum energy restrictions. We have discussed possible origins and
even tantalizing evidence for their existence in the Local Bubble
(\citealt{Gry2014}; \citetalias{MalkMosk_2021}). However, the Epsilon Eridani is particularly
appealing being well-aligned with the direction of the local magnetic
field, yet fitting the basic model restrictions. A serious contradiction
of a large-shock scenario comes from the angular distribution of the
bump particles. It is much easier to reconcile the anisotropy data
with a smaller shock and a narrow propagation channel pertinent to
the Epsilon Eridani. We, therefore, provide in the Appendix a compound acceleration
model tailored for this star in which both the bow-shock and the stellar
wind TS are synergistically involved in the SDA process.


\acknowledgments We thank Samuel Ting, Andrei Kounine and Vitaly
Choutko for numerous discussions of the AMS-02 data, John Krizmanic,
Kazuyoshi Kobayashi, and Shoji Torii for discussions of the CALET
data, Jin Chang and Qiang Yuan for providing the DAMPE data prior
to their publication and discussions, Dmitry Podorozhny, Alexander
Panov, and Andrey Turundaevsky for providing the updated data of NUCLEON
and ATIC, and Eun-Suk Seo for providing the CREAM data. We also acknowledge
numerous discussions with Pat Diamond. Mikhail A. Malkov acknowledges
support from NASA ATP-program within grant 80NSSC17K0255 and from
the National Science Foundation under grants No.\ NSF PHY-1748958
and AST-2109103. Igor V.\ Moskalenko acknowledges support from NASA
Grants No.\ NNX17AB48G, 80NSSC22K0718, 80NSSC22K0477.

\appendix


\section{Compound Acceleration Model\label{subsec:Compound-Acceleration-Model}}

The purpose of this Appendix is to demonstrate that under a favorable
but realistic B-field configuration, the interaction of particles
with the TS is longer than in the standard SDA (shock-drift acceleration)
cycle and results in a larger than the $\Delta p_{\perp}/p_{\perp}\approx\sqrt{r}$
momentum gain, where $r$ is the compression ratio. 
We show that the interaction of star's bow shock with its wind TS in the Epsilon Eridani system and its peculiar location makes it a  
viable candidate source of the observed 10-TV CR bump.

\subsection{Motivation and Physical Context}

A particle acceleration model that would combine a star's bow shock
with its wind TS is motivated by overwhelming evidence
of the Local Bubble origin of 10-TV bump, discussed in the Introduction.
The potential efficiency of this model is supported by studies of
colliding plasma flows in astrophysical \citep[e.g.,][]{Bykov2001}
and laboratory \citep[e.g.,][]{Malkov2022} settings. Here we focus
on a collision region between the stellar wind and the inflowing ISM
wind. 

The absence of traditional CR sources, such as SNRs, in our proximity
raises the bar for potential accelerator efficiency. In Paper I, we
entertained the following three types of possible sources: (i) a radiative
shell of an old SNR or magnetoacoustic shock formed by wave steepening,
(ii) imploding shock within a monolithic Local Bubble model inferred
by \citet{Gry2014} from their observation analysis, and (iii) a stellar
bow-shock. Shocks from (i) and (ii) variety, being comparable in size
with the distance to them, easily provide the required flux of reaccelerated
CRs into the 10-TV range at the observer's location. Indeed, lateral
losses of CRs from the flux tube in Eq.~(\ref{eq:FitForm}) are not
important, given a significant size-to-distance ratio of such source.
By contrast, this ratio is rather small even for a nearby stellar
bow-shock and the lateral spread of reaccelerated CRs might make the
bump unobservable near the Sun. On a potentially positive side, the
bow-shock scenario provides a unique opportunity to associate the
observed CRs with familiar objects, thus increasing the persuasiveness
of this scenario. Adding the serendipitously close (within $6.7^{\circ}$)
alignment of a nearby (ca.~3.2 pc) Epsilon Eridani star with the local
magnetic field direction warrants a closer look at this object as
a likely source of the CR bump.

Our interest in this object is also driven by the angular distribution
of CRs in the rigidity range where the observed CR bump is located.
We have discussed this aspect of the bump phenomenon in \citetalias{MalkMosk_2021} (see
Sect.~7 and Appendix D). We have shown that if the source of the bump
is located on one side of the magnetic field line in a limited range
of 3-10 pc, the observed spectrum will have an anisotropy characterized
by a sharp raise across the magnetic horizon, which is actually observed.
Sources, located at larger distances would have produced a much smoother
angular CR distribution. This is equally true for the rigidity dependence
of the CR spectrum in the bump area, discussed earlier.

\begin{figure}[tb!]
\protect\includegraphics[width=1\linewidth]{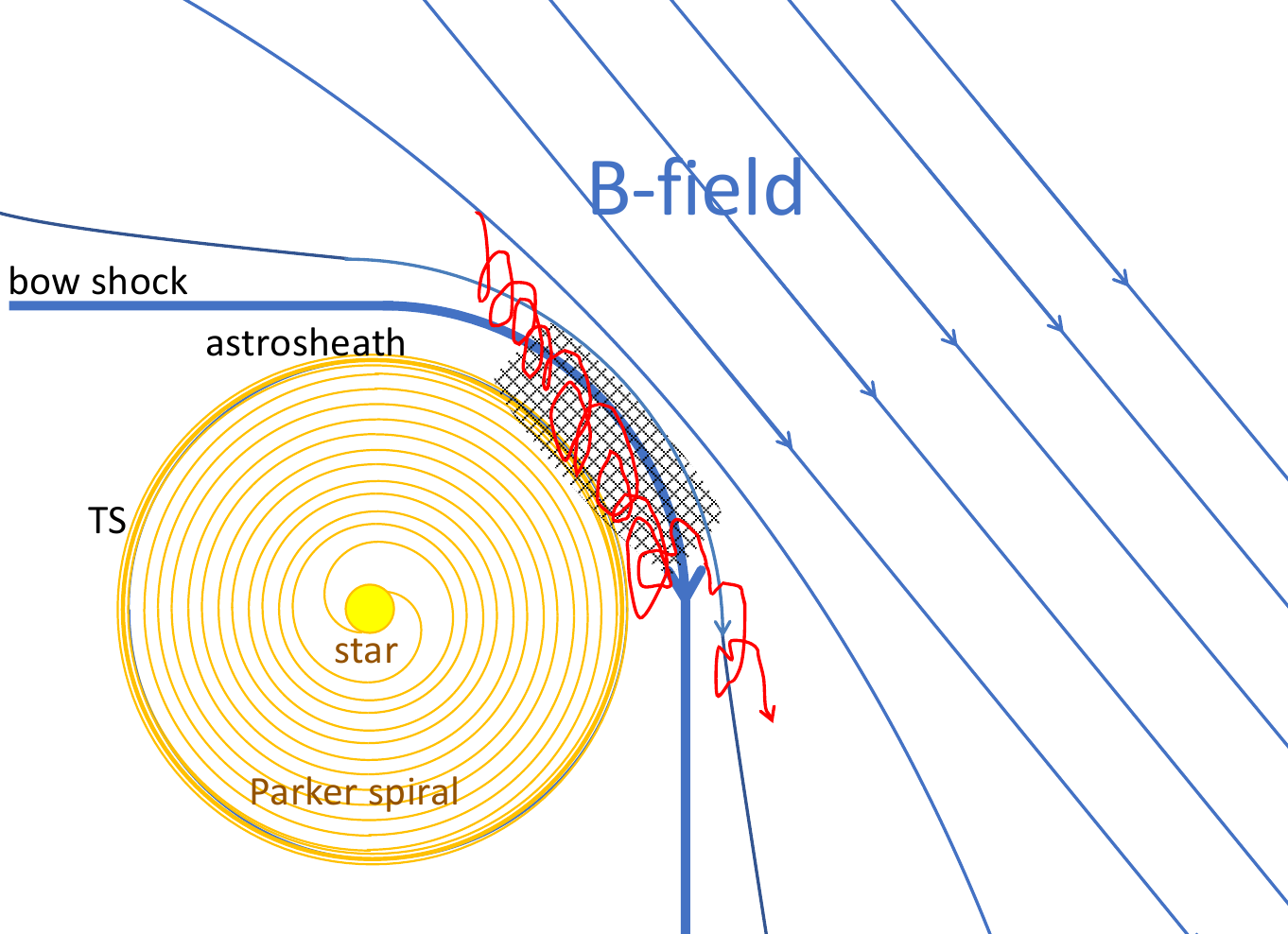}
\protect\caption{Stellar wind cavity showing the TS and bow-shock ahead
of a moving star. The hatched area covering the adjacent portions
of bow-shock and TS is a magnetic trap for CRs (see text).
\label{fig:Stellar-wind-cavity}}
\end{figure}

Estimates of the energy gained by particles after crossing a stellar
bow shock (Sect.~\ref{Sec:discussion}) and the propagation losses
(Sect.~\ref{sec:Reaccel-and-Prop}) demonstrate that the Epsilon Eridani's
bow-shock can marginally account for the bump in terms of its magnitude.
At the same time, additional losses, e.g., those discussed in Paper
I may raise the required CR reacceleration rate at the source. We
therefore consider a possibility of CR reacceleration at the \emph{stellar
wind TS as a booster} for particles that have
already passed through the bow-shock and increased their energy, Fig.~\ref{fig:Stellar-wind-cavity}.

\subsection{Model Description and Estimates of its Efficiency}

Let us first consider plasma conditions in the region between
the stellar wind TS and the bow-shock that includes
the astropause  (that can be characterized as such using its analogy with the
heliopause). This is a turbulent zone of colliding flows where
thermal  particles may undergo stochastic acceleration
\citep[e.g.,][]{Brunetti2007,Lemoine2021}. However, their energy would not be sufficient to reach the solar system, 
so we focus on the reacceleration of preexisting
CRs.

Particles with energies up to tenths of TeV have Larmor radii comparable
with the thickness of the shocked plasma layer between the bow-shock and
TS. Gyrating along this layer, as shown in Fig.~\ref{fig:Stellar-wind-cavity},
they visit the stellar wind cavity where a motion electric field $E\sim U_{\text{SW}}B/c$,
accelerates them. Here $U_{\text{SW}}$ is the velocity of the stellar
wind. As their perpendicular momentum increases by a factor $\Delta p_{\perp}/p_{\perp}\sim U_{\text{SW}}/c$
each time they cross and recross the TS, the pressure of CRs in this
area (shown by a hatched rectangle in Fig.~\ref{fig:Stellar-wind-cavity})
is higher than the ambient CR pressure. By the pressure balance, the
field strength should be lower in this area, thus developing a magnetic
cavity. We may ignore the effect of the ISM wind ram pressure of the
inflowing plasma for now, as the star velocity is relatively low,
$\approx30$ km s$^{-1}$. However, cumulative effects of the magnetic flux
expelled from the cavity, flow compression behind the shocks, involvement
of neutrals, and general flow pattern past the astrosphere result
in a magnetic barrier around the cavity. Therefore, not only the particles
inflowing from the bow-shock apex with $p_{\perp}\gg p_{\parallel}$
become trapped, but also field aligned particles around the cavity
have a good chance to be trapped adiabatically into the magnetic cavity
and stay there, as they are scattered in pitch angle. The adiabatic
trapping into magnetic cavity is quite similar to that into an electrostatic
potential well, as described by \citet{Gurevich68}. We, therefore,
omit the details and focus on the interaction of the trapped particles
with the TS.

\subsection{Dynamics of Trapped Particles}

\begin{figure}[tb!]
\includegraphics[scale=0.49,viewport=30bp 150bp 590bp 770bp]{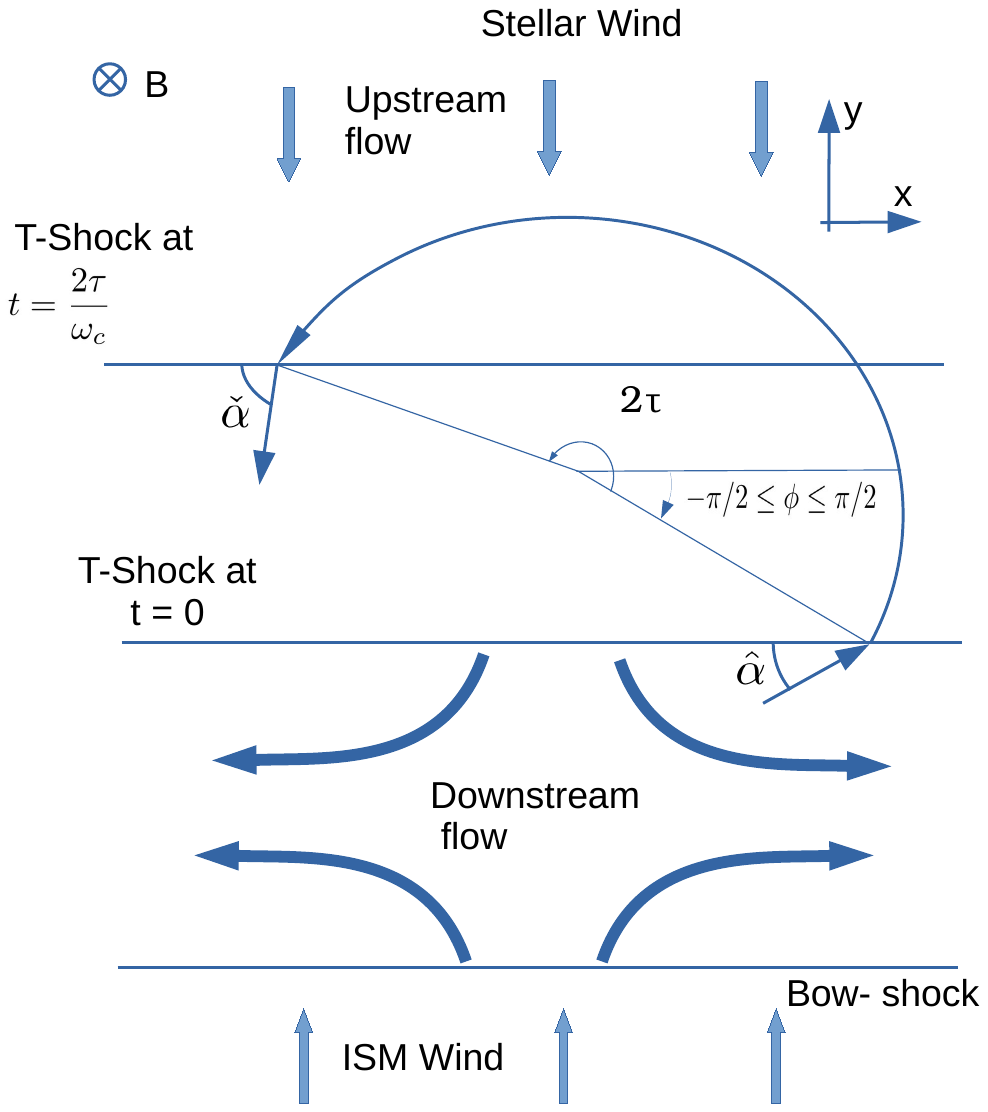}
\caption{Mechanism of acceleration of particles entering the wind cavity from
the downstream of the TS. Particles enter the upstream side of the
TS from the astrosheath, propagate along an arc in the stellar wind
frame, and return to the astrosheath. Positions of the TS 
(T-Shock) at these moments are shown with horizontal lines.\label{fig:TS-BS-acceleration}}
\end{figure}

The purpose of this analysis is to demonstrate that under a favorable
but realistic B-field configuration, the interaction of particles
with the TS is longer than in the standard SDA (shock-drift acceleration)
cycle and, therefore, results in a larger than the $\Delta p_{\perp}/p_{\perp}\approx\sqrt{r}$
momentum gain. It results from the particle orbit crossing by a shock
with the compression ratio $r$. Under the ``favorable'' magnetic
configuration we understand the one where the Parker spiral field
is roughly aligned with
the field between the bow-shock
and TS. Specifically, we assume that the B-field is out of plane on
both sides of  the TS surface, Fig.~\ref{fig:TS-BS-acceleration}. For simplicity,  the curved bow shock and TS are
shown as planar shocks.
Upstream of it, the particle inscribes an arc in the local fluid frame.
The orbit's continuation downstream is more complex, as the motion
electric field is not constant there and cannot be eliminated by the
frame transformation. Indeed, the flow on the downstream side of the
TS is established by its collision with the ISM flow downstream of
the bow-shock. The resulting flow is deflected along the TS-bow-shock
direction away from the flow stagnation region. In and around this
region the flow speed is much lower than $U_{\text{SW}}$ and we may
neglect the effect of the outflow, that is the effect of the motion
electric field on the 10-TV particles. In this case, the particle orbit
is also a part of a circle in the frame fixed by the positions of
the TS and bow shock, and, thereby the magnetic trap.

Assuming that a particle in the magnetic trap enters the wind cavity
through the TS at an angle $\hat{\alpha}$ and returns at angle $\check{\alpha}$,
as shown in Fig.~\ref{fig:TS-BS-acceleration}, we shall obtain the
momentum gain in terms of the entry phase upstream, $\phi:$

\begin{equation}
\phi=-\tan^{-1}\left(\frac{\cos\hat{\alpha}\sqrt{1-U_{\text{SW}}^{2}/c^{2}}}{\sin\hat{\alpha}+U_{\text{SW}}/\hat{v}_{\perp}}\right),\label{eq:fiOf-alpha}
\end{equation}
and the upstream rotation phase, $2\tau$, where $\tau$ is a root
of the following transcendental equation:

\begin{equation}\label{eq:MapTranscen}
\cos\left(\phi+\tau\right)\frac{\sin\tau}{\tau}=\frac{U_{\text{SW}}}{v_{\perp}}.
\end{equation}
Here $v_{\perp}$ is the perpendicular to the B-field component of
particle velocity and $U_{\text{SW}}\ll c$. The equation above can
be derived by assuming that while the particle makes an arc $2\tau$
upstream of the TS the shock itself progresses to a distance $2\tau U_{\text{SW}}/\omega_{c}$ in the upstream frame.
 Note that as long
as particles are in the magnetic trap, their average pitch angle remains
close to $\pi/2$ and we can set $v_{\perp}\approx c$. We represent
the momentum gain as a ratio of the particle Lorentz factors after
and before it visited the wind cavity

\[
\eta=\frac{\check{\gamma}}{\hat{\gamma}}=\frac{1-\left(U_{\text{SW}}/c\right)\cos\left(\phi+2\tau\right)}{1-\left(U_{\text{SW}}/c\right)\cos\phi}.
\]
After reentering the downstream space at the angle $\check{\alpha}$

\begin{equation}
\check{\alpha}=\cos^{-1}\left\{ \frac{\sin\left[\phi+2\tau\right]\sqrt{1-U_{\text{SW}}^{2}/c^{2}}}{1-\left(U_{\text{SW}}/c\right)\cos\left[\phi+2\tau\right]}\right\}, \label{eq:AlphaCheck}
\end{equation}
the particle gyrates around its guiding center fixed in the TS frame,
as we assumed above. Eqs.~(\ref{eq:fiOf-alpha}), (\ref{eq:MapTranscen}),
and (\ref{eq:AlphaCheck}) determine an iterated map $\hat{\alpha}\mapsto\hat{\alpha}^{\prime}$ (Fig.~\ref{fig:Iterated-map-,}).
In the next cycle, the TS entrance angle $\hat{\alpha}^{\prime}=\check{\alpha}$
will be different from $\hat{\alpha},$ in general, but after several
such visits the exit angle $\check{\alpha}$ approaches the entrance
angle $\hat{\alpha}$. In other words, the iterations converge to
a fixed point $\check{\alpha}=\hat{\alpha}=\beta$. This fixed point
is stable, since $\left|d\hat{\alpha}^{\prime}/d\hat{\alpha}\right|<1$
at $\hat{\alpha}=\beta$.

\begin{figure}[tb!]
\includegraphics[width=1\linewidth]{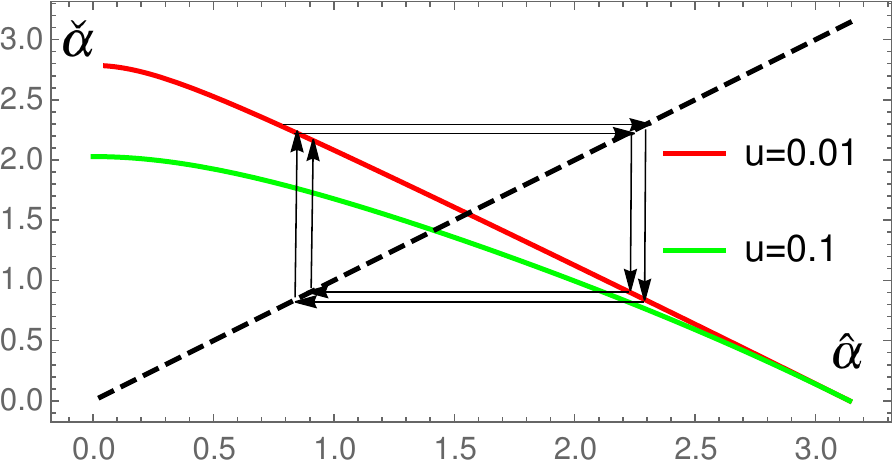}
\caption{Iterated map $\hat{\alpha}\protect\mapsto\hat{\alpha}^{\prime}$.
It is constructed by a superposition of the map $\hat{\alpha}\protect\mapsto\check{\alpha}$
(shown with solid lines for two different stellar wind speed, $u=U_{\text{SW}}/c$)
and the identity map (shown with a dashed line, see text). Arrows
show the first two iterations of an infinite series that ultimately
converges to a fixed point. The case of a faster shock with $u=0.1$
is also shown for comparison with a green line. \label{fig:Iterated-map-,}}
\end{figure}

\subsection{Results and Application to Epsilon Eridani Star}

Since the stellar wind is nonrelativistic, $U_{\text{SW}}\ll c$,
from Eqs.~(\ref{eq:fiOf-alpha}) and (\ref{eq:MapTranscen}) we find
$\tau\approx\pi\left(1-U_{\text{SW}}^{2}\right)/2\approx\pi/2$ and
$\phi\approx\pi U_{\text{SW}}/2$. Under these conditions, the energy
gain tends to $\eta\approx1+2U_{\text{SW}}/c$. It corresponds to
the energy gain of a particle specularly reflected from the medium
moving at the speed $U_{\text{SW}}$, as expected. The question that
arises is for how long a particle can remain in the magnetic trap
and keep gaining momentum $\Delta p_{\perp}/p_{\perp}=2U_{\text{SW}}/c$
after each rotation.

To give a simple answer to this question, we consider the energy range
wherein the above acceleration mechanism operates. The upper bound
is set by the particle gyroradius not to exceed the size of magnetic
trap. Using the results of \citet{Wood2002}, we may assume that the
size of the trap across the field is $\sim$$10^{3}$AU. As the field
outside of the magnetic cavity should be enhanced according to the
above considerations and the Axford-Crandfield effect \citep{1972NASSP.308..609A}
to about 10 $\mu$G, particles of several tenths of TV can be confined
to the cavity and accelerated, even though the field should be diminished
inside.

Using the available data about the Epsilon Eridani star, we may also
estimate the stellar wind velocity $U_{\text{SW}}$ that determines
the particle acceleration rate. The required parameters are the mass
loss rate $\dot{M}$ which is about $30\dot{M}_{\sun}$ and the stand-off
distance of TS being $R_{\text{SW}}\sim10^{3}$AU at its minimum on
the upwind side of the moving star \citep{Wood2002}. Since these
are very crude estimates, we simply assume that the Epsilon Eridani
is moving through the ISM, similar to that around our heliosphere,
that is with the same pressure and the ISM wind having nearly the
same speed. Therefore, the ram pressure at the TS is also nearly the
same: $\rho_{\text{SW}}U_{\text{SW}}^{2}=\rho_{\text{\ensuremath{\sun}}}U_{\sun}^{2}$.
From the mass ratio 30, we obtain $\rho_{\text{SW}}U_{\text{SW}}R_{\text{SW}}^{2}=30\rho_{\text{\ensuremath{\sun}}}U_{\sun}R_{\sun}^{2}$,
where the index SW refers to the Epsilon Eridani. We, therefore, obtain
$U_{\text{SW}}/U_{\sun}\simeq R_{\text{SW}}^{2}/30R_{\sun}^{2}\simeq3$
and find that the momentum gain per cycle is $\Delta p_{\perp}/p_{\perp}=2U_{\text{SW}}/c\sim10^{-2}$.
We thus conclude that particles trapped in the magnetic cavity for
a hundred of gyro-periods gain an additional energy comparable with
that they gained by crossing the stellar bow shock.

\bibliographystyle{aasjournal}
\bibliography{}


\end{document}